\documentclass[11pt]{article}
\usepackage{latexsym}
\usepackage{amssymb}
\usepackage{amsmath}
\usepackage[hypertex]{hyperref}
\usepackage{graphicx}
\usepackage{axodraw}

\newcommand{\Mpl}{M_{\rm Pl}}
\newcommand{\ra}{\rightarrow}
\newcommand{\iso}[2]{\,^{#1}\mathrm{#2}}
\newcommand{\lsim}{\lesssim}

\begin{document}

\pagestyle{empty}

\begin{flushright}
SLAC-PUB-11795, 
hep-ph/0604027 \\
\end{flushright}

\vspace{1.1cm}

\begin{center}
\mbox{\bf\LARGE A Universe Without Weak Interactions}

\vspace*{1.3cm}
{\large Roni Harnik$^{1}$, Graham D. Kribs$^{2}$, and Gilad Perez$^{3}$} \\
\vspace*{1.3cm}

\mbox{$^1$\textit{Stanford Linear Accelerator Center, 
Stanford University, Stanford, CA 94309 and}} \\[1mm]
\mbox{\textit{Physics Department, Stanford University, Stanford, CA 94305}}
\\[2mm]

\mbox{$^2$\textit{Department of Physics and 
Institute of Theoretical Science}} \\
\mbox{\textit{University of Oregon, Eugene, OR 97403}} \\[2mm]

\mbox{$^3$\textit{Theoretical Physics Group, 
Ernest Orlando Lawrence Berkeley National Laboratory,}} \\
\mbox{\textit{University of California, Berkeley, CA 94720}} \\[2mm]

\vspace*{1cm}

\texttt{roni@slac.stanford.edu},~ 
\texttt{kribs@uoregon.edu},~ 
\texttt{gperez@lbl.gov}

\end{center}

\vspace*{0.2cm}

\begin{abstract}

A universe without weak interactions is constructed that undergoes 
big-bang nucleosynthesis, matter domination, 
structure formation, and star formation.  The stars in this universe 
are able to burn for billions of years, synthesize elements 
up to iron, and undergo supernova explosions, dispersing 
heavy elements into the interstellar medium.
These definitive claims are supported by a detailed analysis where
this hypothetical ``Weakless Universe'' is matched to our 
Universe by simultaneously adjusting Standard Model and 
cosmological parameters.  For instance, chemistry and nuclear physics 
are essentially unchanged.
The apparent habitability of the Weakless Universe suggests that
the anthropic principle does not determine the scale of electroweak 
breaking, or even require that it be smaller than the Planck scale,
so long as technically natural parameters may be suitably adjusted.
Whether the multi-parameter adjustment is realized or probable 
is dependent on the ultraviolet completion, such as the string
landscape.  
Considering a similar
analysis for the cosmological constant, however, 
we argue that no adjustments of other parameters are able to
allow the cosmological constant to raise up even remotely close 
to the Planck scale while obtaining macroscopic structure.
The fine-tuning problems associated with the electroweak breaking scale 
and the cosmological constant therefore appear to be qualitatively 
different from the perspective of obtaining a habitable universe.  

\end{abstract} 

\newpage
\pagestyle{plain}

\section{Introduction}
\label{introduction-sec}

Naturalness suggests that the weak interaction ought to be far weaker 
than observed in our Universe.  In this paper we calculate the
gross features of a universe without weak interactions.
Specifically, we show that by adjusting Standard Model and
cosmological parameters we are able obtain a universe that is 
remarkably similar to our own.  
This ``Weakless Universe''\footnote{Throughout this paper 
we refer to the hypothetical universe without weak interactions
as the ``Weakless Universe'', and the universe that the reader
and the authors inhabit as ``our Universe''.} has 
big-bang nucleosynthesis, structure formation, star formation, 
stellar burning with 
a wide range of timescales, stellar nucleosynthesis up to iron and 
slightly beyond, and mechanisms to disperse heavy elements through 
type Ia supernovae and stellar mergers.

Examining the consequences of parameter adjustments on our Universe
has a long history (see Refs.~\cite{Carr:1979sg,BarrowTipler} 
for some early discussion).  The typical argument is as follows:  
hold all parameters fixed except for one, vary this one,
and find how much it can be changed before the universe with
the parameter adjustment appears qualitatively different than 
without the adjustment.  Often ``disasters'' occur, where
the modified universe looks radically different and apparently
unsuitable for observers.  This suggests that there is an 
anthropic rationale for parameters to take on values that 
are consistent with observations in our Universe.

This logic was applied to the cosmological constant (CC) by Weinberg
\cite{Weinberg:1987dv}.  
The CC is by far the most troubling parameter for effective field
theorists since (1) it is the most technically unnatural parameter,
i.e., it must be tuned to one part in $10^{120}$, 
and (2) we have no understanding why it should be numerically
small but non-zero.
As the CC is increased, holding everything else fixed, 
the universe eventually becomes CC dominated early enough
that density perturbations in the universe do not have enough
time to grow and go non-linear to form large scale structure.
Applying a sensible probability distribution to the value of the CC,
Weinberg argued that the CC should be near to the bound set by the
anthropic rationale.  Remarkably, astrophysical observations are
now consistent with a CC that is within about two orders of magnitude
of Weinberg's anthropic prediction.  
This may be viewed as support for varying just the CC in an
ensemble of universes.  

Interestingly, even if other astrophysical parameters are varied, 
such as the amplitude of density perturbations or the baryon asymmetry, 
the bound on the CC is not significantly relaxed beyond the original 
Weinberg bound \cite{Tegmark:1997in,Tegmark:2005dy}
(but see also \cite{Aguirre:2001zx}).  
This analysis assumes these parameters can be freely varied, but the 
prior probability distribution of these parameters may be sensitive 
to the details of inflation 
\cite{Graesser:2004ng,Feldstein:2005bm,Hall:2006ff}.
Nonetheless, the analysis of \cite{Tegmark:2005dy} may serve
as moral justification for varying several low energy parameters 
simultaneously.

The other technically unnatural parameter, the electroweak scale,
also appears to be anthropically limited to a narrow range
\cite{Agrawal:1997gf,Jeltema:1999na,Oberhummer:2000mn,Hogan:2006xa}
when \emph{only} the electroweak scale is varied.  
We do not disagree with this conclusion.  Our viewpoint, however,
is orthogonal to these papers.  We wish to ask the following question:
Is there any universe without 
electroweak interactions that looks anything like our Universe?
Here ``any universe'' refers to allow ourselves to arbitrarily 
adjust Standard Model and cosmological parameters.  This amounts
to adjusting technically natural parameters (since all of
the other Standard Model parameters are technically natural,
and it is easy to imagine specific models or mechanisms in which 
cosmological parameters are also determined by technically 
natural parameters).  We should emphasize that there is nothing 
``wrong'' with simultaneously adjusting technically natural parameters 
of an effective field theory!  It is only when the Standard Model 
is embedded into a precise ``ultraviolet completion'' when such
adjustments may or may not be warranted.

What we will show is that there is no need to anthropically rationalize 
the small size of the electroweak breaking scale so long as 
Standard Model and cosmological parameters can be adjusted.
This is not a trivial argument, since we require ``matching'' 
the Weakless Universe to our Universe including, 
for instance, stellar fusion with long-lifetime stars.  This 
requirement appears immediately problematic since the reaction 
$pp \ra D + e^+ + \nu_e$ is absent.  
We show how this and other apparent problems can be overcome through 
a careful analysis following suitable parameter adjustments.
In the end, the readers must judge for themselves whether the outcome 
of our exercise provides a sufficiently interesting universe.  

We do not engage in discussion of the likelihood of doing simultaneous
tunings of parameters nor the outcome of statistical ensembles of parameters.
These questions are left up to the ultraviolet completion,
such as the string landscape, which is outside of the scope of 
effective field theory.
Instead, we are interested in ``running the universe forward'' 
from a time after inflation and baryogenesis through billions of years 
of evolution.  We will exploit the knowledge of our Universe 
as far as possible, adjusting Standard Model and cosmological parameters 
so that the \emph{relevant} micro- and macro-physical outcomes match 
as closely as possible.  We emphasize that this is really a practical 
matter, not one of principle, since any significant relaxation of the 
``follow our Universe'' program would be faced with horrendously 
complicated calculations.  Put another way, there is probably a 
wide range of habitable universes with parameters and structures 
that look nothing like our Universe.
For us, it is enough to find one habitable
Weakless Universe about which we can make the most concrete statements, 
hence matching to our Universe as closely as possible.

We define a habitable universe as one having big-bang nucleosynthesis,
large-scale structure, 
star formation, stellar burning through fusion for long lifetimes
(billions of years) and plausible means to generate and disperse heavy
elements into the interstellar medium.  As a consequence, we will
demand ordinary chemistry and basic nuclear physics be largely unchanged 
from our Universe so that matching to our Universe is as straightforward
as possible.  
We are not aware of insurmountable obstacles extending our analysis 
to planet formation, habitable planets, and the generation of 
carbon-based life.  Nevertheless, these questions are beyond the scope 
of this paper, and we do not consider them further.

Finally, we should emphasize from the outset that this paper 
represents a purely theoretical exercise.  There are no 
(even in principle) experimental predictions.  The purpose of 
our paper is to provide a specific, concrete counter-example 
to anthropic selection of a small electroweak breaking scale.

\section{The Weakless Universe}
\label{proposal-sec}

The Weakless Universe is described by a set of parameters that 
we will relate to the SM and cosmological parameters of our 
observed Universe.  In subsequent sections, we will explain
these parameter choices in detail, and describe the consequences.

The Weakless Universe contains the gauge group $SU(3)_c \times U(1)_{\rm em}$. 
This is equivalent to introducing 
$SU(2)_L \times U(1)_Y$ with a Higgs mechanism in which the
Higgs (mass)$^2$ $\ra \Mpl^2$.  In this sense, the Weakless Universe
allows the Higgs (mass)$^2$ to take on its most natural value.  
The strengths of color, electromagnetic, and gravitational forces 
are fixed by setting
\begin{eqnarray}
\tilde{\alpha}_{\rm em}(Q^2 = 0) &=& \alpha_{\rm em}(Q^2 = 0) \\
\tilde{\Lambda}_{\rm QCD} &=& \Lambda_{\rm QCD} \\
\tilde{M}_{\rm Pl} &=& \Mpl
\end{eqnarray}
where tildes correspond to the parameters of the Weakless Universe.
We choose to match the scale of confinement, $\Lambda_{\rm QCD}$,
but this is of course equivalent to matching the strength of the 
strong force in a perturbative regime $Q^2 \gg \Lambda^2_{\rm QCD}$.
As we will see below, the number of fermion species in the Weakless 
Universe is different than in our Universe, and thus both 
the electromagnetic and strong couplings have mild differences in 
the logarithmic running of their gauge couplings toward the UV\@.

Given just the color and electromagnetic forces in the low energy 
effective theory of the Weakless Universe, fermions are vector-like.
Thus, there are no constraints on the fermion content at low energies 
due to anomaly-cancellation.  For instance, there is no need for
fermions to come in complete generations.  Moreover, fermion masses 
are gauge-invariant without a Higgs mechanism and can be set to
whatever values we like.  The Weakless Universe contains the 
up, down, and strange quarks as well as the electron with masses
\begin{eqnarray}
\tilde{m}_f &=& m_f \qquad f = u,d,s,e \; .
\end{eqnarray}
The other fermions in our Universe are not present
(the reasoning behind keeping the strange quark will be discussed
in much more detail in Sec.~\ref{QCD-sec}).
Of course all of the fermion masses given here are equivalent
to having $SU(2)_L \times U(1)_{\rm em}$ broken by a Higgs mechanism 
with $\tilde{v} \ra \Mpl$ and a large (technically natural) hierarchy 
between Yukawa couplings of the light species 
$\lambda_{u,d,s,e} \sim 10^{-22 \ra -19}$ and those of the 
heavy species $\lambda_{c,b,t,\mu,\tau} \sim 1$.\footnote{Neutrinos
also become heavy in this limit given dimension-5 operators 
$(L H)^2/\Mpl$ with order one coefficients.}
The coincidence that fermion masses and $\tilde{\Lambda}_{\rm QCD}$
are within a few orders of magnitude in mass in the Weakless Universe 
is the rephrasing of the coincidence that the electroweak scale 
is within three orders of magnitude of $\Lambda_{\rm QCD}$
in our Universe.

The unbroken global symmetries of the Weakless Universe consist 
of $U(1)_u \times U(1)_d \times U(1)_s \times U(1)_e$, 
generalized from baryon and lepton number of our 
Universe.\footnote{Higher dimensional operators suppressed by
the Planck scale violate these individual fermion numbers in the 
Weakless Universe just like baryon and lepton number in our Universe.}
Hence, individual quark number and electron number, not just baryon and 
lepton number, are conserved.  This has significant implications
for both the origin of the initial abundances of fermions 
as well as the stability of bound states of quarks.  
Quark number is ultimately confined into baryons,
as we explain in Sec.~\ref{freezeout-sec}.  
The fermion number densities themselves are determined by 
``fermiogenesis'' (generalized from baryogenesis) 
as will be explained in Sec.~\ref{fermiogenesis-sec}.

There are several cosmological parameters that are set in the
Weakless Universe as follows.
First, the total energy density is given by
\begin{eqnarray}
\tilde{\Omega}_{\rm total} &=& \Omega_{\rm total} \; = \; 1 \; ,
\end{eqnarray}
which assumes that the Weakless Universe is flat.
We take the flatness, as well as scale-invariant density 
perturbations with an amplitude $\delta \rho/\rho \sim 10^{-5}$,
to arise from inflation, just like our Universe.  
The total matter density 
\begin{eqnarray}
\tilde{\Omega}_{\rm matter} &\simeq& \Omega_{\rm matter} \; \simeq \; 0.23 \; .
\end{eqnarray}
The difference $\Omega_{\rm total} - \Omega_{\rm matter}$ is
taken to be vacuum energy, the same as our Universe. 

The matter density of the Weakless Universe is fixed to be the 
same as the matter density of our Universe, so that the 
transition from radiation domination to matter domination occurs 
at the same epoch in both Universes.  
Matter consists of visible baryons (protons and neutrons)
and dark matter, analogous to our Universe.  Dark matter could 
consist of free neutral hyperons or a non-baryonic candidate 
(or both) as we discuss in Sec.~\ref{darkmatter-sec}.

At BBN the visible matter can be described by two parameters,
the ratio of the visible baryon abundance to photons 
$\tilde{\eta}_b$ and the ratio of protons to neutron abundance.
We take 
\begin{eqnarray}
\tilde{\eta}_b &\simeq& 4 \times 10^{-12} \; \simeq \; 10^{-2} \, 
\eta_b \; ,
\end{eqnarray}
where we emphasize that this corresponds to just the
baryon asymmetry in protons and neutrons, not hyperons.
This is taken to be about two orders of magnitude smaller than 
in our Universe.  This judicious parameter adjustment allows 
the Weakless Universe 
to have a hydrogen-to-helium ratio the same as our Universe 
without strong sensitivity to the ratio of the proton to neutron 
abundance.
Hence, the first galaxies and stars are formed of roughly the 
same material as in our Universe.  
Moreover, the lower helium abundance that results from the lower
baryon asymmetry occurs simultaneous with a substantially increased 
abundance of deuterium.  The much increased deuterium abundance allows 
stars in the Weakless Universe to ignite through proton-deuterium fusion,
explained in detail in Sec.~\ref{stellarnuc-sec}.

Given a smaller ratio of visible matter to dark matter, some aspects 
of large scale structure are affected including the maximum visible
mass of galaxies as well as the formation of disks.  We will briefly 
mention these effects in Sec.~\ref{perturbations-sec},
citing previous results that indicate no disasters occur 
if the visible baryon abundance in the Weakless Universe is two 
orders of magnitude smaller than in our Universe.

\section{QCD}
\label{QCD-sec}

In the Weakless Universe, QCD gets strong and confines at 
$\tilde{\Lambda}_{\rm QCD} = \Lambda_{\rm QCD}$, the same as 
in our Universe.  The number of light(er) quarks is the same as our
Universe ($u$, $d$, $s$) so we expect the low energy properties
of QCD to be unchanged.  In particular, several well-known sum rules 
in QCD relate these fundamental parameters to the parameters of 
the chiral Lagrangian, such as the Gell-Mann--Oakes--Renner 
relation
\begin{eqnarray}
m_{\pi}^2 f_{\pi}^2 &=& (m_u + m_d) \Lambda_{\rm QCD}^3 \; ,
\end{eqnarray}
which implies the product 
$\tilde{m}_{\pi}^2 \tilde{f}_{\pi}^2 = m_{\pi}^2 f_{\pi}^2$
is the same across both universes.
This is not surprising since we can construct the low-energy
effective Lagrangian involving pions and photons that
has the same physics in the Weakless Universe as it does in
our Universe.  Differences in pion masses or
couplings between the Weakless Universe and our Universe arise 
only due to radiative corrections associated with heavier 
particles and the weak interactions, but these effects are
extremely small.  

Baryon masses, however, are more subtle.  The difference 
between the proton and neutron mass is determined by the difference in
quark masses as well as the electromagnetic splitting.  The
mass difference in the Weakless Universe is therefore the same 
as our Universe.
The absolute value of nucleon masses, however, receive a large 
contribution to their mass from gluons as well as an important 
smaller contribution from strange quarks.  The mass fraction 
attributed to the strange content can be estimated using the
approximate SU(3) flavor symmetry among the light quarks applied to 
hyperon masses.  For example, Ref.~\cite{DynamicsSM} find that
\begin{eqnarray}
\frac{\langle N | m_s \overline{s} s | N \rangle}{2 m_{N}} &\sim&
130 \ra 260 \; \mbox{MeV}
\end{eqnarray}
where the uncertainty depends on whether or not higher order chiral
corrections are included.  This implies that between
about 15\% to 30\% of nucleon mass arises as a result of the 
strangeness content of the nucleon.  While the size of this result 
is not without controversy (see for example \cite{Kaplan:1989fc}), 
it is clear that the strange quark does lead to a contribution 
to the nucleon mass that amounts to between a few to ten or more percent
of its mass.  This is sufficient to have very significant effects 
on basic nuclear processes.  For instance, deuterium is bound by only 
2 MeV which is far smaller than these estimates of the strange contribution
to the nucleon mass.

This is the origin of why we hold the strange quark mass fixed in the
Weakless Universe.  By retaining strange in the low energy effective
theory, nucleon masses in the Weakless Universe are the same as in 
our Universe to very high accuracy (up to effects associated 
with the charm or heavier quark content of the hadrons which is 
expected to be an extremely small effect).  Sensitive nuclear properties 
including deuterium binding, the triple-$\alpha$ process, etc., are 
unaffected.  Moreover, given the same baryon masses, pion masses, 
and couplings, all nuclear physics associated with the residual strong 
force between nucleons in the Weakless Universe is the same as our 
observed Universe to extremely high accuracy.

There is, of course, one important difference in the Weakless Universe:
hadrons are stable against weak decay, and so free neutrons as well as
pions and several strange hadrons are stable.  Unraveling the
effects of having several stable hadrons in addition to the proton 
will occupy the next few sections.

\section{Freeze Out Abundances of Hadrons}
\label{freezeout-sec}

Light mesons, which carry an individual fermion (quark) number, 
such as kaons and charged pions, are stable in the Weakless Universe.
Those without quark number, such as the $\pi^0$ and $\eta$, 
are unstable and can decay to photons through the axial anomaly,
as in our Universe.
Given that there are several stable hadrons in the Weakless Universe,
it is important to understand the distribution of the matter abundance 
in stable hadrons after the QCD phase transition but before~BBN\@.

The result, as we explain below, is that even though the light mesons 
and the heavier strange baryons (such as the excited states with 
higher isospin)
cannot decay, at finite temperature their densities are 
highly suppressed.  Thus, to a very good approximation, the hadronic 
content of the Weakless Universe just before BBN consists of only protons, 
neutrons and the lightest stable strange hadron, the $\Lambda_s^0$ hyperon.
Using the chiral Lagrangian it straightforward to see that
light mesons and baryons are in thermal equilibrium down to
temperatures of order BBN\@.  Thermal equilibrium is maintained through
rapid $t$- or $u$-channel strong interaction processes involving the 
exchange of light hadrons.

First we show how the universe is driven to a configuration with no
light charged mesons.
Suppose that at temperatures well below the QCD phase transition a
sizable fraction of the up density consisted of $\pi^+$ (as well as
protons and neutrons).  Available processes include the interaction of
$\pi^+$ with a neutron that convert into a proton and a $\pi^0$ through
neutron exchange.  Clearly individual quark number is conserved by this
process.  Then since the neutral pion can quickly decay 
(at that epoch the corresponding decay rate is much faster than the 
expansion rate of the Universe), the number density of neutral pions
stays at roughly the thermal density, $n_{\pi^0}\sim\exp\left(-m_\pi/T\right)$.
But the thermal density of neutral pions is tiny, so that the inverse 
process is exponentially suppressed, and so effectively all the charged 
pions transfer their quark number to baryons in the thermal bath.  
Similar considerations apply to the charged and neutral kaons once 
the neutron is replaced with $\Lambda_s^0$ and $\pi^0$ with an $\eta$ meson.

Now suppose that a sizable density of $\Sigma^+$ baryons is present in
the plasma.  Processes which involve $\pi^+$ $t$-channel exchange will
induce $\Sigma^+ + n \to \Lambda_s^0 + p$ transitions.  The
mass difference, $\Delta m$, between the outgoing and incoming
particles is much larger than the temperature, $T_{\rm BBN}$, 
just before BBN starts 
\begin{eqnarray}
\frac{\Delta m}{T_{\rm BBN}}
   &\simeq& \frac{m_{\Sigma^+}-m_{\Lambda_s^0}}{T_{\rm BBN}} \; \sim \; 80 \; .
\end{eqnarray}
Similar considerations also apply for the $\Sigma^{0,-}$.
Thus at $T_{\rm BBN}$ the net excess of strange density is equal to the 
density of $\Lambda_s^0$ to a very good approximation.  To summarize, 
apart from a sizable fraction of $\Lambda_s^0$ baryons proportional to
the number density of strange quarks, the abundance of stable free 
mesons and other heavy stable baryons is highly suppressed.

Given that virtually all of the strange quarks are confined into the 
$\Lambda_s^0$ hyperon, we now estimate hyperon binding with protons 
and neutrons.  
Protons and neutrons bind together only into deuterium,
and with a rather small binding energy, $2.2$ MeV\@.  
The binding energy arises through 
charged and neutral single pion exchange as well as four-fermion
operators suppressed by the cutoff scale of the chiral Lagrangian.  
The $\Lambda_s^0$, however, cannot exchange of a single pion with a 
proton or neutron since $\Lambda_s^0$ is an isospin-singlet.  
Exchange a single $I=1/2$ kaon is possible, but this is highly suppressed 
by the Yukawa potential 
of exchanging such a massive scalar particle across the characteristic 
inverse distance associated with nuclear binding, of order a few to 
tens of MeV\@.  Another process is the exchange of two pions in a 
one-loop diagram, but the loop must also contain an isospin-triplet 
baryon, namely 
one of the $\Sigma^{\pm,0}$ hyperons that are 80 MeV heavier than the 
$\Lambda_s^0$.  
These considerations suggest that the $\Lambda_s^0$ has \textit{far} 
weaker strong interactions with other baryons.

This particular issue has been studied in some detail in our Universe
in the context of the search for and study of ``hyper-nuclei'', i.e., 
nuclei with one or more hyperons bound to the nucleus~\cite{Gibson:1995an}.
Since weak interactions lead to a relatively slow decay rate, 
bound states with hyperons can be produced and studied experimentally.
Also, theoretical work has been applied to hyper-nuclei binding, using
various potential models as well as lattice QCD to estimate binding 
energies.  The result of these studies 
is that ``hyper-deuterium'' (deuterium with the neutron substituted for
a $\Lambda_s^0$) has neither been observed experimentally nor is it 
expected to be a bound state, consistent with our qualitative analysis 
above.  The lightest experimentally observed 
bound state with a hyperon is ``hyper-tritium'', namely a bound 
state with one proton, one neutron, and one $\Lambda_s^0$.  This 
is extremely weakly bound with a binding energy of only $130$~keV 
more than that of deuterium.
Contrast this with ordinary tritium, which 
is bound by about 6 MeV more than deuterium.  The absence of a bound state of
hyper-deuterium combined with the rather weak binding of 
hyper-tritium suggests that the $\Lambda_s^0$ plays essentially 
no role in affecting BBN predictions of the very lightest elements
so long as the strange quark number density is not orders of magnitude
larger than the up and down quark number density.  The $\Lambda_s^0$
hyperons are therefore expected to become a form of neutral, stable, 
baryonic dark matter in the Weakless Universe.

\section{Fermiogenesis}
\label{fermiogenesis-sec}

The Weakless Universe is assumed to have vanishing total 
electric charge, 
\begin{eqnarray}
\tilde{Q}_{\rm Universe} &=& Q_{\rm Universe} \simeq 0 \; ,
\label{electrically-neutral-eq} 
\end{eqnarray}
which relates the number density of quarks with electrons,
\begin{eqnarray}
2 n_u - n_d - n_s &=& 3 n_e \; .
\label{number-densities-eq}
\end{eqnarray}
The visible baryon asymmetry at BBN can be related to a 
different function of the quark number densities
\begin{eqnarray}
\tilde{\eta}_b &=& \frac{1}{3} \frac{n_u + n_d - 2 n_s}{n_\gamma}
\label{etab-def}
\end{eqnarray}
where the $-2 n_s$ accounts for removing hyperons.  
Given Eqs.~(\ref{number-densities-eq}) and (\ref{etab-def}),  
two parameters remain which we can take to be the strange number
density $n_s$ and the ratio 
\begin{eqnarray}
\left. \frac{n_n}{n_p} \right|_{\rm eff} &=& 
\frac{n_u + 2 n_d - 3 n_s}{2 n_u + n_d - 3 n_s} \; .
\label{np-frac}
\end{eqnarray}

We showed in Sec.~\ref{freezeout-sec} that quark number 
is ultimately confined into baryons, thus quark number 
densities can indeed be translated into baryon number densities.  
For number densities of up, down, and strange that are roughly 
comparable, we expect $n_{\Lambda_s^0} = n_s$,
and the ratio of the remaining up and down quark number densities becomes 
the ratio of protons to neutrons.  These estimates certainly change
if there are huge disparities between one quark species to another.  
For example, if $n_s \gg n_u,n_d$ then the huge excess of strange quarks 
would have no choice but to bind together into $\Omega^-$ hyperons.
Similarly, if $n_u = n_d = n_s$ equal to high accuracy, then 
$n_e \simeq 0$ by Eq.~(\ref{number-densities-eq}) and the Universe 
would be composed of just $\Lambda_s^0$ hyperons.  Clearly the 
relative number densities of quarks plays an essential role in 
determining the properties of a Weakless Universe.

Like baryon or lepton number in our Universe, these 
individual quark numbers are free parameters.  Unlike our
Universe, however, individual quark flavor and lepton flavor is
conserved.  This means that it is not sufficient to simply have
a mechanism of baryogenesis, since this relies on weak interactions
to distribute the baryon number to the light quark species.
Moreover, leptogenesis is also inoperative in the Weakless Universe,
since electroweak sphalerons converting lepton number into baryon number
are absent.

These considerations imply that baryogenesis must be generalized
to fermiogenesis:  the origin of the fermion numbers of the 
Weakless Universe.  Of course we do not know the mechanism of
baryogenesis in our Universe any more than we know what to match it
to in the Weakless Universe.  The simplest proposal is to just
assume particular initial conditions for the quark numbers 
(specifically, $n_s$ and the ratio $n_u/n_d$), which is tantamount 
to assuming an excess $B-L$ number in our Universe 
as the origin of baryons.  

Alternatively, mechanisms that do not rely on sphalerons could 
be employed, such as spontaneous baryogenesis \cite{Cohen:1988kt}
or gravitational baryogenesis \cite{Davoudiasl:2004gf}.
These mechanisms require a source of baryon number 
violation in our Universe, and hence fermion number violation
in the Weakless Universe.  It is straightforward to construct
models with higher dimensional operators that can populate
the fermion numbers as we see fit (with no new technically 
unnatural parameters) since the flavor structure of these 
operators is arbitrary.  We will not speculate about this in detail, 
but suffice to say there is nothing particularly unusual about 
assuming $n_u \sim n_d \sim \mathcal{O}({\rm few}) \times n_s$ 
given a relatively flavor democratic fermiogenesis mechanism.
In this case, 
the strange number goes into $\Lambda_s^0$ and the remaining
roughly equal up and down numbers go into roughly equal numbers
of protons and neutrons.  We will show that this setup
(as well as order one variations in the quark number densities)
leads to a perfectly habitable Weakless Universe.

\section{BBN}
\label{BBN-sec}

Big-bang nucleosynthesis
amounts to a complicated interplay between nucleons binding
together through strong interactions in a bath of a huge number 
photons capable of breaking them up.  
Given $n_u \sim n_d$, the Weakless Universe enters the BBN phase 
with approximately equal numbers of protons and neutrons causing 
several interesting and important differences on the primordial 
element abundances.  If the visible baryon-to-photon ratio 
$\tilde{\eta}_b$ 
were the same as in our Universe, then it is easy to estimate that
nearly all protons and neutrons get absorbed into helium.  
A helium-dominated Universe is qualitatively different than
our Universe, from the formation of disks in galaxies,
to star formation, to the nuclear burning cycle.  This may well
be an interesting Weakless Universe, but it is not one with which
we can reliably calculate.  Hence, if we wish to follow our 
Universe as closely as possible, it is clear
that this outcome must be changed.  

There are two possibilities.  One is to change the ratio of
protons and neutrons by adjusting $n_u \not= n_d$.
For instance, setting $n_u \simeq 5 n_d/3$, which corresponds to
$n_p/n_n = 7$, roughly reproduces
the proton-to-helium ratio in our Universe for the
same baryon asymmetry.
A second possibility is to leave the number densities of up
and down comparable, and instead adjust the visible baryon asymmetry 
(the total number density of quarks) to be about two orders
of magnitude lower than in our Universe.
The latter is the route we will take, for reasons that
will be clear by the end of this section.

Numerical calculations are 
needed to make more precise estimates of BBN\@.  We have modified
the standard BBN code \cite{BBN} to compute the elemental abundances
in the Weakless Universe.  There are four main differences
when computing BBN in the Weakless Universe:
First, of course, all weak interactions are decoupled and the neutron 
is stable.  Second, the initial conditions entering the BBN phase
depend only on the initial number densities of hadrons that themselves
are determined by the number densities of quarks.  We will illustrate
what happens for a set of three possibilities:  
$n_p = n_n \gg n_{\Lambda_s^0}$;
$n_p = 9 n_n \gg n_{\Lambda_s^0}$; and 
$n_p = n_n/9 \gg n_{\Lambda_s^0}$
[These three cases equivalent to $n_u = (1, 19/11, 11/19) n_d$ 
using Eq.~(\ref{np-frac})].
In all cases we have neglected the strange quark number density 
since we already
showed that at best it could only weakly bind with deuterium to
form hyper-tritium.  But since tritium itself is a subdominant
component (both in our Universe and in the Weakless Universe)
hyper-tritium is even more irrelevant due to its far smaller
binding energy.  Third, we neglect the effects of having stable
mesons around, which is an excellent approximation as we explained above.
Finally, there are no neutrinos in the Weakless Universe, and so
the number of relativistic degrees of freedom during BBN is
slightly smaller causing a slightly slower expansion rate;
this leads to a very modest effect on the results, but we 
nevertheless took it into account. 

\begin{figure}[t]
\centerline{\includegraphics[width=0.85\hsize]{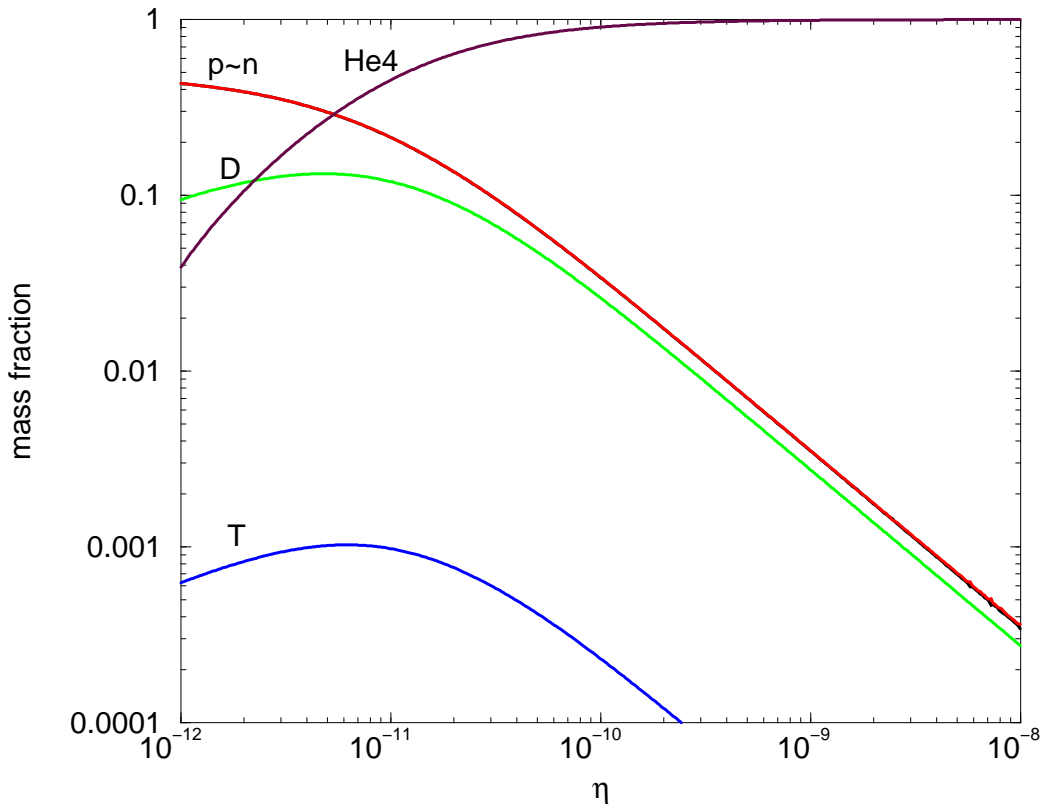}}
\caption{Elemental abundance mass fractions as a function of
the visible baryon-to-photon ratio assuming $n_p = n_n \gg n_{\Lambda_s^0}$.}
\label{bbn1-fig}
\end{figure}

\begin{figure}[t]
\centerline{\includegraphics[width=0.85\hsize]{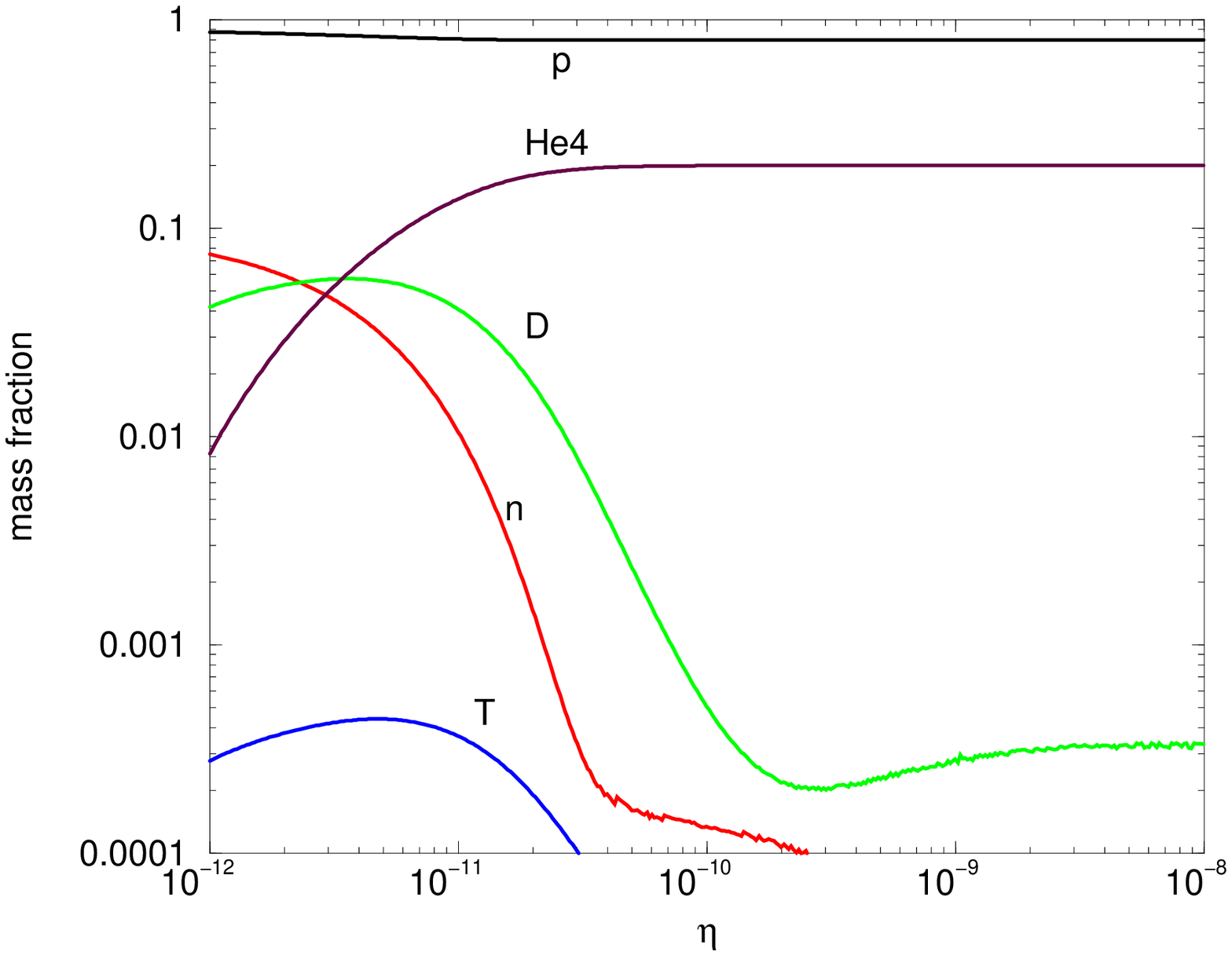}}
\caption{Same as Fig.~\ref{bbn1-fig} except $n_p = 9 n_n \gg n_{\Lambda_s^0}$.}
\label{bbn2-fig}
\end{figure}

\begin{figure}[t]
\centerline{\includegraphics[width=0.85\hsize]{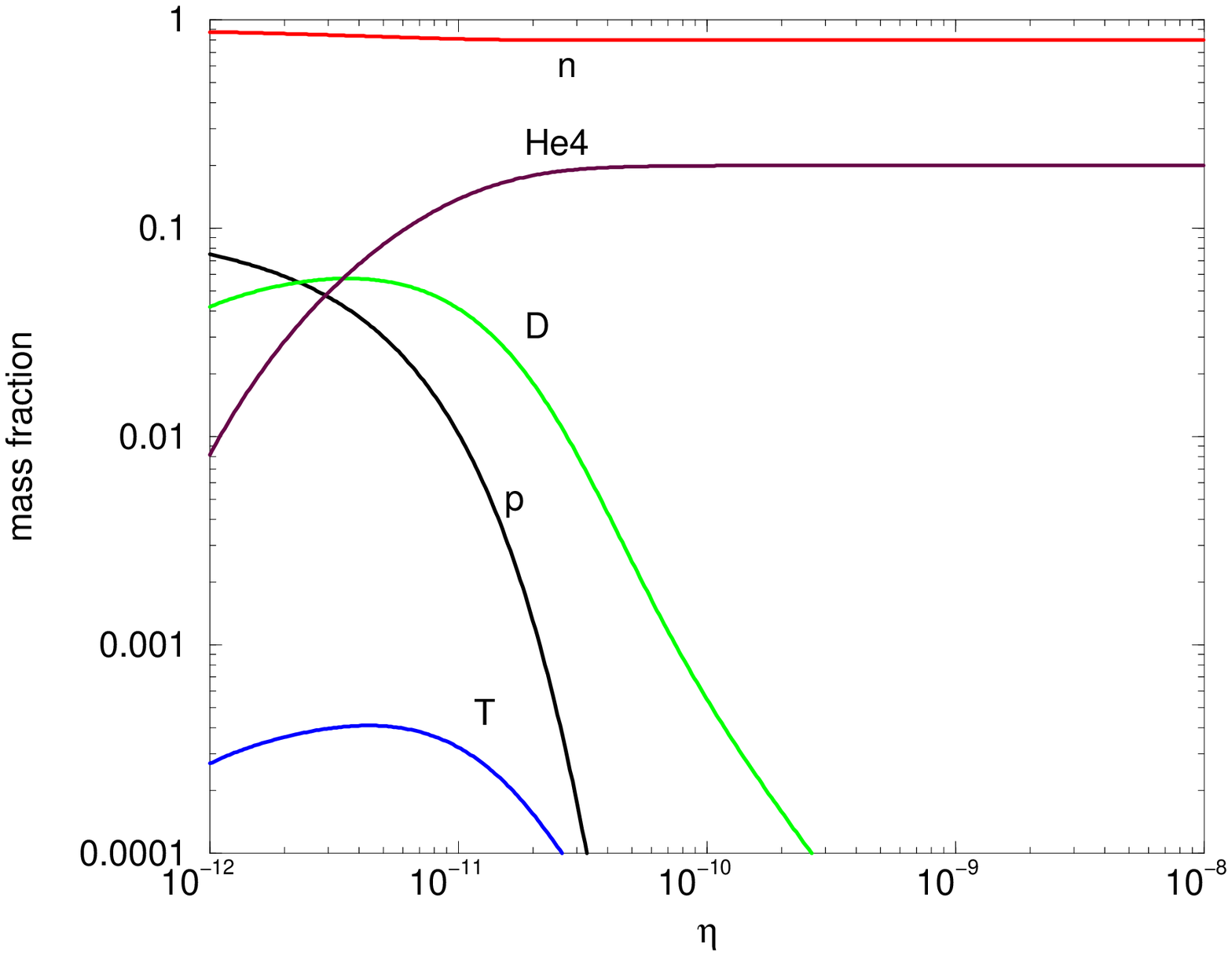}}
\caption{Same as Fig.~\ref{bbn1-fig} except $n_p = n_n/9 \gg n_{\Lambda_s^0}$.}
\label{bbn3-fig}
\end{figure}

Given these assumptions, the result of our numerical simulations of
BBN are shown in Figs.~\ref{bbn1-fig},\ref{bbn2-fig},\ref{bbn3-fig}.
The visible baryon-to-photon ratio was varied over a wide range 
for illustration.  
Exactly what baryon abundance is optimal for a Weakless Universe
depends on initial conditions and on the outcome that we seek.

These calculations lead us to one fascinating result:  if the visible
baryon asymmetry of the Weakless Universe is lowered to roughly 
$\tilde{\eta}_b \simeq 4 \times 10^{-12}$, a substantial
fraction of the synthesized nuclei is deuterium, of order 10\% by mass.  
This conclusion is not particularly sensitive to the initial
number densities of quarks in the range $n_u/9 < n_d < 9 n_u$,
as the figures show.  This can be qualitatively understood
in that by reducing the baryon asymmetry 
we are delaying the initiation of net deuterium production, 
allowing deuterium to be produced at a lower temperature. 
The rate of deuterium fusion into $^4$He is reduced due to the
Coulomb barrier, leaving a high abundance of unfused deuterium.

As we will discuss in Sec.~\ref{stellarnuc-sec} below, 
the existence of a substantial 
fraction of deuterium provides a mechanism to ignite stars through 
proton-deuterium fusion, avoiding the proton-proton reaction that
is absent in the Weakless Universe.  This is the reason that we
take the baryon asymmetry to be $\tilde{\eta}_b \simeq 4 \times 10^{-12}$
rather than adjusting the number densities of up and down quarks.
For this baryon asymmetry and $n_u \sim n_d$, the resulting 
hydrogen-to-helium mass fraction is equivalent to our Universe, 
roughly $Y_H/Y_{He} \simeq 3$.
Nevertheless, a far larger fraction of hydrogen is deuterium 
rather than single protons, so that a hydrogen gas cloud has a 
composition of about 1/4 deuterium and 3/4 protons by mass.
Note also that there is generally a substantial free neutron abundance
for $\tilde{\eta}_b \simeq 4 \times 10^{-12}$.  These electrically 
neutral relics will behave as form of baryonic dark matter
but with a small abundance $\Omega_n \lsim 10^{-4}$.

\section{Chemistry}
\label{chem-sec}

In the post-BBN phase of the Universe, the main players in our
Universe are electromagnetism and gravity.  Both of the these
forces are unchanged in the Weakless Universe.  The elemental abundances 
of the Weakless Universe have also been matched to our Universe 
(and are chemically indistinguishable, aside from the irrelevant 
tiny abundance of lithium).  Chemistry in the Weakless Universe 
is virtually indistinguishable from that of our Universe.  
The only differences are the higher fraction of deuterium as hydrogen
and the absence of atomic parity-violating interactions.  

Maintaining this similarity between the Universes relies on having
only one stable charged lepton:  the electron.  
The presence of muons or taus 
(with masses as observed in our Universe) would allow for various 
exotic chemical properties and nuclear reaction rates.  For instance, 
the Coulomb barrier would be far smaller for atoms with orbiting 
muons or taus, allowing dense-packed molecules and fusion at extremely 
low temperatures.  Though this could be an interesting 
universe\footnote{Consider a universe with $\tilde{m}_\tau = m_\tau$
with baryons dominantly in $\iso{4}{He}$.  A $\iso{4}{He}$ nucleus 
with one $\tau^-$ and one electron will behave chemically like
hydrogen.  This is because the double-charged nucleus 
will be screened to single-changed by the tight orbital of the
$\tau^-$.  This provides a potential mechanism to allow $\iso{4}{He}$
fusion at much lower temperatures as well as forming hydrogen-like
molecules that are important for star formation, 
see Sec.~\ref{stars-sec}.}
it does not match our Universe and so we choose to remove muons and taus 
from the Weakless Universe.  

Other more benign effects occur if the heavier quarks ($c,b,t$) 
were present in the Weakless Universe.  Given that individual 
quark number is conserved, the lightest baryons carrying a heavy quark 
are stable.  This means in addition to protons, neutrons, and 
$\Lambda_s^0$ hyperons, there would be several new stable baryons
including $\Lambda_c^+$, $\Lambda_b^0$ and 
$\Lambda_t^+$.\footnote{Several species of isospin non-singlets 
are also stable, but their primordial abundance is expected to be 
far smaller than the isospin-singlets for the same reason 
that the $\Sigma^{\pm,0}$ abundance is far smaller than the
$\Lambda_s^0$ as discussed above.}
If these exotic stable baryons were in significant abundance in 
the Weakless Universe, there would be numerous anomalously heavy 
isotopes of hydrogen (and heavier elements).  These are not obviously 
an impediment to successful BBN or star formation, but it would change 
the details of stellar nucleosynthesis reactions in ways that we 
are not able to easily calculate.  Again, following our program of 
matching to our Universe as closely as possible, we eliminate this 
problem by insisting that the Weakless Universe is devoid of these
heavy quarks.

\section{Matter Domination and the Growth of Density Perturbations}
\label{perturbations-sec}

One extremely important epoch in our Universe is the era of 
matter domination.  This is when density perturbations are able 
to grow linearly and subsequently collapse to form structure.  
This must occur before 
any structure-destroying epoch takes place, such as if a cosmological 
constant eventually dominates the energy density.  
Matching to our Universe suggests that we take 
$\tilde{\Omega}_{\rm total} = \Omega_{\rm total} = 1$ and
$\tilde{\Omega}_{\rm matter} = \Omega_{\rm matter} \simeq 0.23$.
There is certainly no need for a cosmological constant to 
dominate the latter epoch of the Weakless Universe; instead it could be 
an open universe with $\tilde{\Omega}_{\rm total} \simeq 0.23$,
or dark matter could make up the difference between a closed
universe and the baryons and radiation.  Both of these
possibilities, however, are expected to have ``higher order'' effects 
on the evolution of our Universe (for instance, a larger dark matter 
abundance implies matter-domination occurs at a slightly earlier epoch).
Matching to our Universe as closely as possible suggests that 
we take the Weakless Universe to have the cosmological parameters 
as given originally in Sec.~\ref{proposal-sec}.

Since we have chosen to hold the $n_H/n_{He}$ fraction constant
with $\tilde{\eta}_b \sim 4 \times 10^{-12}$, the visible baryon density
in the Weakless Universe is similarly reduced, 
$\Omega_{\rm visible \; baryons} \sim 10^{-4}$.
The smaller visible baryon abundance implies we should expect
an overall decrease in the number density of stars in galaxies,
although calculating precisely the scaling relies on knowing 
detailed dynamics of galaxy formation.

The maximum size of a given large scale structure to form in the
universe depends on the scale of the initial perturbation
and the time with which it can grow.  Ref.~\cite{Hellerman:2005yi} 
found that the maximum mass $M^{\rm max}$ that any structure 
which can form in the universe can be expressed simply as
\begin{eqnarray}
M^{\rm max} &\lsim& \Mpl^3 \delta_0^{3/2} \rho_\Lambda^{-1/2}
\label{hell-eq}
\end{eqnarray}
where $\delta_0$ is the size of the density perturbation
and $\rho_\Lambda$ is the size of the cosmological constant density.
Putting in values for our Universe, one finds
$M^{\rm max} \lsim 10^{16} M_\odot$.  Our galaxy contains
about $10^{11} M_\odot$ baryons, which implies one can form
a lower bound on the ratio of visible matter to dark matter 
of about $10^{-5}$ so that galaxies like our own are formed.  
The Weakless Universe with $\tilde{\Omega}_{\rm visible \; baryons}/
\tilde{\Omega}_{\rm dark \; matter} \sim 10^{-3}$
clearly satisfies this bound.

Once structures of galactic size are able to form, 
the gas of baryons must be able to cool sufficiently into
a rotationally supported disk from which stars can be created.
Ref.~\cite{Tegmark:2005dy} found that there is a 
more stringent lower bound on the ratio of visible matter to
dark matter from this constraint.  The precise numerical
value of the lower bound depends on the mechanism of cooling 
(hydrogen line cooling or molecular hydrogen cooling) and the 
virial temperature of the galaxy or structure.  Given these
uncertainties, they found the lower bound to be of order 
$3 \times 10^{-3} \ra 10^{-4}$.  Ref.~\cite{Tegmark:2005dy} 
also showed that molecular 
viscosity can relax this lower bound by another order of magnitude, 
so a fair conclusion from this analysis is that one begins
to run into trouble only when the baryon to matter density
ratio falls below about $10^{-4}$.  Again, the Weakless Universe
satisfies this bound.

\section{Dark Matter Candidates}
\label{darkmatter-sec}

We have already remarked that free neutrons and free $\Lambda_s^0$ hyperons 
would behave as baryonic dark matter.  This is amusing, since the 
abundance of (baryonic) dark matter becomes closely connected with 
the matter abundance in the Universe.  However, given that the 
total baryon abundance is two orders of magnitude smaller 
in the Weakless Universe, this dark matter is insufficient to
cause the Universe to become matter-dominated at a time
the same as (or close to) our Universe.  The Weakless
Universe therefore needs non-baryonic dark matter to match
to our Universe.

The usual candidates for non-baryonic dark matter contemplated for 
our Universe are not necessarily effective in the Weakless
Universe.  Weakly interacting massive particles (WIMPs) with an
electroweak scale mass whose abundance is determined by the
usual thermal freeze-out after electroweak decoupling are 
obviously unsuitable since electroweak interactions are absent!
(This does not preclude a non-thermal source of $100$ GeV mass
mass particles, but the mass scale here is clearly arbitrary.)
It is straightforward to add a dark matter sector that is 
unconnected with weak interactions, such as axions or primordial
black holes.  This is the working assumption that we have in
mind for the Weakless Universe.  

Yet another possibility is to deviate from the parameters of the
Weakless Universe in the following way.  
Take $\tilde{\Omega}_{\rm baryons} = \tilde{\Omega}_{\rm matter}$ 
with $n_u = n_d = 1.0001 n_s$.
Then the vast majority (99.99\%) of quarks are confined 
into $\Lambda_s^0$ hyperons, which act as baryonic dark matter. 
The reminder go into ``visible'' protons and neutrons with an 
abundance that remains of order $10^{-4}$ so that BBN largely 
follows Fig.~\ref{bbn1-fig}.  This Universe is harder to calculate,
since such a large abundance of $\Lambda_s^0$ hyperons during
BBN may well lead to a more significant abundance of exotic
elements including hyper-tritium and hyper-helium.  While we 
expect this would be a perfectly viable alternative, the modifications
to BBN and stellar nucleosynthesis make it harder for us to match
to our Universe and so we will not consider this amusing possibility 
any further.

\section{Heavy Element Isotopic Stability}
\label{isotope-sec}

In our Universe, only a narrow band in the atomic mass-number 
($A$--$Z$) plane contain 
stable elements.  In the Weakless Universe, the band is much wider, 
extending out to the edge of the ``valley of stability'' 
well known from nuclear physics.  In particular, since both 
$\beta$-capture and $\beta$-decay are absent, a much wider
range of proton-rich and neutron-rich isotopes are stable.  
Using tables from Ref.~\cite{isotope}, we can estimate the 
isotopic stability table for the Weakless Universe by relabeling
all previously unstable (to $\beta$-capture or $\beta$-decay) 
isotopes as being stable.  
This is shown in Fig.~\ref{isotope-fig} for elements up to oxygen.  
\begin{figure}
\begin{picture}(440,300)
  \Line(60,0)(90,0)
  \Line(30,30)(210,30)
  \Line(30,60)(240,60)
  \Line(60,90)(270,90)
  \Line(120,120)(300,120)
  \Line(120,150)(330,150)
  \Line(120,180)(360,180)
  \Line(120,210)(390,210)
  \Line(180,240)(420,240)
  \Line(180,270)(420,270)
  \Line(30,30)(30,60)
  \Line(60,0)(60,90)
  \Line(90,0)(90,90)
  \Line(120,30)(120,210)
  \Line(150,30)(150,210)    \DashLine(210,30)(210,60){4}
  \Line(180,30)(180,270)    \DashLine(240,60)(240,90){4}
  \Line(210,60)(210,270)    \DashLine(270,90)(270,120){4}
  \Line(240,90)(240,270)    \DashLine(300,120)(300,150){4}
  \Line(270,120)(270,270)   \DashLine(330,150)(330,180){4}
  \Line(300,150)(300,270)   \DashLine(360,180)(360,210){4}
  \Line(330,180)(330,270)   \DashLine(390,210)(390,240){4}
  \Line(360,210)(360,270)   \DashLine(420,240)(420,270){4}
  \Line(390,240)(390,270)
  \DashLine(210,30)(225,30){3}
  \DashLine(240,60)(255,60){3}
  \DashLine(270,90)(285,90){3}
  \DashLine(300,120)(315,120){3}
  \DashLine(330,150)(345,150){3}
  \DashLine(360,180)(375,180){3}
  \DashLine(390,210)(405,210){3}
  \DashLine(420,240)(435,240){3}
  \DashLine(420,270)(435,270){3}
  \Line(270,270)(270,285)
  \Line(300,270)(300,285)
  \Line(330,270)(330,285)
  \Line(360,270)(360,285)
  \Line(390,270)(390,285)
  \Line(420,270)(420,285)
  \Text(75,15)[c]{$n$}
  \Text(45,45)[c]{$\iso{1}{H}$}
  \Text(75,45)[c]{$\iso{2}{H}$}
  \Text(105,45)[c]{$\iso{3}{H}$}
  \Text(135,45)[c]{$\iso{4}{H}$}
  \Text(165,45)[c]{$\iso{5}{H}$}
  \Text(195,45)[c]{$\iso{6}{H}$}
  \Text(75,75)[c]{$\iso{3}{He}$}
  \Text(105,75)[c]{$\iso{4}{He}$}
  \Text(135,75)[c]{$\iso{5}{He}$}
    \Line(120,60)(150,90)
    \Line(120,90)(150,60)
  \Text(165,75)[c]{$\iso{6}{He}$}
  \Text(195,75)[c]{$\iso{7}{He}$}
    \Line(180,60)(210,90)
    \Line(180,90)(210,60)
  \Text(225,75)[c]{$\iso{8}{He}$}
  \Text(135,105)[c]{$\iso{6}{Li}$}
  \Text(165,105)[c]{$\iso{7}{Li}$}
  \Text(195,105)[c]{$\iso{8}{Li}$}
  \Text(225,105)[c]{$\iso{9}{Li}$}
  \Text(255,105)[c]{$\iso{10}{Li}$}
    \Line(240,90)(270,120)
    \Line(240,120)(270,90)
  \Text(135,135)[c]{$\iso{7}{Be}$}
  \Text(165,135)[c]{$\iso{8}{Be}$}
    \Line(150,120)(180,150)
    \Line(150,150)(180,120)
  \Text(195,135)[c]{$\iso{9}{Be}$}
  \Text(225,135)[c]{$\iso{10}{Be}$}
  \Text(255,135)[c]{$\iso{11}{Be}$}
  \Text(285,135)[c]{$\iso{12}{Be}$}
  \Text(135,165)[c]{$\iso{8}{B}$}
  \Text(165,165)[c]{$\iso{9}{B}$}
    \Line(150,150)(180,180)
    \Line(150,180)(180,150)
  \Text(195,165)[c]{$\iso{10}{B}$}
  \Text(225,165)[c]{$\iso{11}{B}$}
  \Text(255,165)[c]{$\iso{12}{B}$}
  \Text(285,165)[c]{$\iso{13}{B}$}
  \Text(315,165)[c]{$\iso{14}{B}$}
  \Text(135,195)[c]{$\iso{9}{C}$}
  \Text(165,195)[c]{$\iso{10}{C}$}
  \Text(195,195)[c]{$\iso{11}{C}$}
  \Text(225,195)[c]{$\iso{12}{C}$}
  \Text(255,195)[c]{$\iso{13}{C}$}
  \Text(285,195)[c]{$\iso{14}{C}$}
  \Text(315,195)[c]{$\iso{15}{C}$}
  \Text(345,195)[c]{$\iso{16}{C}$}
  \Text(195,225)[c]{$\iso{12}{N}$}
  \Text(225,225)[c]{$\iso{13}{N}$}
  \Text(255,225)[c]{$\iso{14}{N}$}
  \Text(285,225)[c]{$\iso{15}{N}$}
  \Text(315,225)[c]{$\iso{16}{N}$}
  \Text(345,225)[c]{$\iso{17}{N}$}
  \Text(375,225)[c]{$\iso{18}{N}$}
  \Text(195,255)[c]{$\iso{13}{O}$}
  \Text(225,255)[c]{$\iso{14}{O}$}
  \Text(255,255)[c]{$\iso{15}{O}$}
  \Text(285,255)[c]{$\iso{16}{O}$}
  \Text(315,255)[c]{$\iso{17}{O}$}
  \Text(345,255)[c]{$\iso{18}{O}$}
  \Text(375,255)[c]{$\iso{19}{O}$}
  \Text(405,255)[c]{$\iso{20}{O}$}
\end{picture}
\caption{The expected isotopic stability table in the Weakless Universe.
Only isotopes with $A \le 20$ and $Z \le 8$ are shown.  Reliable 
knowledge of neutron-rich isotopes is not extensive for most 
low $Z$ elements; those not shown on the right-hand side are not 
necessarily unstable to neutron emission.  The proton-rich isotopes 
not shown on the left-hand side are unstable to proton emission.  
The crossed-out elements are unstable to proton, neutron, or 
$\alpha$ emission.}
\label{isotope-fig}
\end{figure}
Elements which are unstable to weak decay are presumably stable
against proton, neutron, or $\alpha$ emission.\footnote{However, we cannot
rule out tiny mass splittings that might allow decay via strong
interactions that happened to be accidentally much weaker than
weak interactions due to kinematic suppression.}

The conservation of both proton number and neutron number, 
combined with the smaller total number of neutrons over protons 
in a primordial gas cloud implies that proton-rich 
isotopes are likely to be populated.  Recall, the primordial gas cloud 
is expected to contain 75\% hydrogen and 25\% helium, in which about 
1/4 of the hydrogen is deuterium by mass.  None of these primordial 
elements has more neutrons than protons.  Combine this initial 
condition with the fact that isotopes with an equal or excess 
number of protons over neutrons never spontaneously emit a neutron 
implies that stellar nucleosynthesis is expected to create elements 
along the $A \le 2 Z$ side of the valley of stability.  

The critical point in which the last element with equal numbers of protons 
and neutrons ($A=2 Z$) is stable in the Weakless Universe 
against decay into a $Z-1$ element and a proton is 
$\iso{76}{Sr}$.\footnote{Actually, there is one higher isotope that 
is stable, $\iso{84}{Mo}$, but it is separated from $\iso{76}{Sr}$
by gaps at $A = 78,80,82$ whereas all of the $A = 2 Z$
elements at and below $\iso{76}{Sr}$ are stable.}
This is well above iron, suggesting that stellar fusion is 
perfectly capable of creating nearly the entire first half
of the periodic table.
In our Universe, by contrast, the last stable element on the 
$A = 2 Z$ line is $\iso{40}{Ca}$.  
To produce elements above $\iso{76}{Sr}$ in the Weakless Universe
requires stellar nucleosynthesis reactions that emit neutrons.  
In our Universe, one known candidate reaction is 
$\iso{22}{Ne} + \alpha \ra \iso{24}{Mg} + n$ (that would occur in the 
late stage of neon burning) leading to a ``$s$-process'' to 
synthesize heavier elements.  This requires, however, weak interactions 
to get to the neutron-rich side of neon ($A_{\rm Ne} > 20$).  While we 
cannot rule out the possibility of other reactions that emit neutrons 
in the later stages of stellar burning, they appear to be unlikely 
since they would follow a path away from the bottom of the valley 
of stability.  It appears unlikely to produce any significant 
abundance of elements above about $Z = 38$ (Sr).  

In our Universe, superheavy elements well above iron are created 
in supernovae.  In a later section we will discuss supernovae 
in the Weakless Universe, but clearly the conservation of proton 
and neutron number, combined with the absence of weak decays, 
prevents the usual pathways to creating superheavy elements.  
There are amusing consequences of the absence of heavy elements 
and radioactive decay.  For instance, the existence of a molten core 
that is continually heated by radioactive uranium and thorium in 
earth-like planets would \emph{not} occur in the Weakless Universe.  
Plate tectonics, volcanos, geothermal heat, etc., would not exist 
billions of years after planets form.  Nevertheless, we do not view 
this difference with our Universe as anything more than 
a curiosity.

\section{Star Formation}
\label{stars-sec}

In order for a dilute gas to collapse into a bound object it 
must lose energy, satisfying the virial theorem.  Heuristically, 
one imagines that an atom with some initial kinetic energy falls
into a potential well.  The only way for the atom to be bound 
into the potential without climbing out is for it to lose energy 
as it falls inside.  The first condition for star formation is thus 
an efficient cooling mechanism whereby the gas can radiate energy 
out of the system. 

There are several mechanisms for a gas cloud to lose energy.  
An efficient mechanism to cool a gas cloud down to temperatures 
of order $T \sim 1$ eV is line radiation from primordial hydrogen 
and helium.  This corresponds to a Jeans mass 
\begin{eqnarray}
M_J &\simeq& 10^6 \left( \frac{T}{1 \; {\rm eV}} \right)^{3/2} 
\left( \frac{n}{10^4 \; {\rm cm}^3} \right)^{-1/2} M_\odot \; .
\end{eqnarray}
Clearly, to get anywhere close to stellar mass objects, it is crucial
to lower the temperature of the gas even further.
The density of the gas, $n$, must also increase, but this cannot
continue arbitrarily, since the gas cloud is expected to form a 
rotationally supported disk.

To lower the gas temperature below what hydrogen or helium 
line radiation can accomplish requires either other gas components
(elements created from the first stars, i.e., a sub-dominant 
metallicity component with line radiation at lower energies) 
or another mechanism.  The first stars would not have other
gas components, and it is generally believed that molecular
hydrogen plays an important role to cool the gas to of order
$T \sim 0.01$ eV through soft vibrational modes.  Molecular hydrogen 
is typically formed through gas-electron collisions where free electrons 
act as catalysts.  In Ref.~\cite{Oh:2001ex} it was shown that a 
non-equilibrium abundance of molecular hydrogen is expected to form 
in the gas clouds of the early Universe that lead to the first stars.

This is the ultimate origin of our requirement that the outcome of 
BBN lead to a chemical composition in the Weakless Universe that is 
similar to our Universe.  For instance, if baryons were
buried dominantly in helium, such as would exist for a baryon-to-photon
ratio equal to our Universe (with $n_u \simeq n_d$), 
no molecular transitions would be available to cool the gas below
what helium line radiation could accomplish since helium does not
form molecules.  Sub-dominant components may well allow for helium
to condense.  For instance, with 
$\tilde{\eta}_b = \eta_b \simeq 5 \times 10^{-10}$ and
$n_u = n_d \gg n_s$, we find that BBN results in about 97\%
by mass in helium and about 2\% by mass in hydrogen.  
Hence, there is a sub-dominant component that could allow for
cooling below $T \sim \,\, {\rm eV}$.
However, helium itself does not begin to fuse until the temperature
of the gas is much higher than for hydrogen fusion, and this
ultimately leads to far fewer stars that live only as long as
helium burning permits, which is tens of millions of years.
Again, this may well lead to an habitable universe, 
but it is not one with which we can make reliable predictions
due to the difficulties of simulating helium gas cloud collapse,
ignition, etc.

\section{Stellar Nucleosynthesis}
\label{stellarnuc-sec}

In our Universe, a collapsing gas cloud with hydrogen begins to fuse
by the weak interaction
\begin{eqnarray}
p + p &\rightarrow& \mbox{D} + e^+ + \nu_e \; 
\label{pp-eq}
\end{eqnarray}
once the temperature of the core rises above about $10^7$ K\@.
This reaction is obviously absent in the Weakless Universe
and naively appears to be serious bottleneck to stellar ignition.
However, we showed that for $\tilde{\eta}_b \simeq 4 \times 10^{-12}$
and $n_u \simeq n_d$, the synthesized elements of BBN 
included a large fraction of deuterium.

The presence of a much higher fraction of deuterium allows collapsing 
clouds of hydrogen gas to skip the slow, weak interaction mediated 
process (\ref{pp-eq}) and go directly to
\begin{eqnarray}
\label{D+p}
p + \mbox{D} &\rightarrow& \iso{3}{He} + \gamma \; . \label{magic-eq}
\end{eqnarray}
Subsequent burning occurs by the standard chains
\begin{eqnarray}
\iso{3}{He} + \iso{3}{He} &\rightarrow& \iso{4}{He} + 2 p \label{He3-eq} \\
   \mbox{D} + \iso{3}{He} &\rightarrow& \iso{4}{He} + p \; ,
\end{eqnarray}
to generate energy through fusion up to $\iso{4}{He}$.
The hydrogen-to-helium fusion chain in the Weakless Universe is 
therefore effectively
\begin{eqnarray}
2 p + 2 \mbox{D} &\rightarrow& \iso{4}{He} + 2 p 
\end{eqnarray}
where the protons act as catalysts to fuse deuterium into
$\iso{4}{He}$.  The deuterium burning reaction is very fast 
and thus can change the dynamics of stellar burning in the 
Weakless Universe.  Below we remark on how this leads to 
modest changes in the stellar mass-lifetime relation.

In our Universe, burning above helium is ``blocked'' by 
the absence of stable elements at atomic mass numbers 
$A=5$ and $A=8$.
In the Weakless Universe, the $A=5$ gap persists, namely 
$\iso{5}{Li}$ is unstable to proton emission and $\iso{5}{He}$ 
is unstable to neutron emission (see Fig.~\ref{isotope-fig}).  
The $A=8$ gap partially persists given that $\iso{8}{Be}$ is unstable 
to breakup into two $\alpha$'s, but $\iso{8}{B}$, $\iso{8}{Li}$, and 
$\iso{8}{He}$ are all stable.  Of these, $\iso{8}{Li}$ and $\iso{8}{He}$ 
are unlikely to play any role.  This is because the neutron-rich side 
of the valley of stability is virtually impossible to reach at this point 
in stellar nucleosynthesis since it requires starting with neutron-rich 
ingredients, of which there are none.\footnote{There is a 
small but non-trivial mass fraction of tritium, of order
$0.1\%$ [see Fig.~\ref{bbn1-fig}].  However, 
the $p + T \ra \iso{4}{He} + \gamma$ reaction is by far 
the fastest reaction for tritium burning, and thus we expect this 
neutron-rich isotope to long since have disappeared into 
$\iso{4}{He}$.}

Once the stellar core temperature gets high enough, the helium 
can begin to fuse via the famous triple-$\alpha$ process
\begin{eqnarray}
3 \left( \iso{4}{He} \right) \; \rightarrow \;
\iso{8}{B} + \iso{4}{He} &\rightarrow& \iso{12}{C} + \gamma
\end{eqnarray}
where a non-equilibrium abundance of $\iso{8}{B}$ is built up 
and consumed into $\iso{12}{C}$.
Then, carbon can be pushed further up the nucleotide table
through reactions such as
\begin{eqnarray}
\iso{12}{C} + \iso{4}{He} &\rightarrow& \iso{16}{O} + \gamma \; ,
\end{eqnarray}
that occur our Universe, as well as new ``proton clumping'' reactions 
such as
\begin{eqnarray}
\iso{12}{C} + p &\rightarrow& \iso{13}{N} + \gamma \\
\iso{13}{N} + p &\rightarrow& \iso{14}{O} + \gamma \; .
\end{eqnarray}
In the Weakless Universe, $\iso{13}{N}$ and $\iso{14}{O}$ 
are stable (see Fig.~\ref{isotope-fig}) but $\iso{15}{F}$ is not, 
and so this particular proton clumping trajectory stops at oxygen.

New reaction chains to fuse up to proton-rich isotopes are possible.
For example, in the Weakless Universe there are trajectories that 
lead to $\iso{9}{C}$, $\iso{10}{C}$, and $\iso{11}{C}$ which are
all stable.  The starting point to enter the proton-rich side is 
$\iso{3}{He}$.  Naively one might guess the reaction
\begin{eqnarray}
\iso{3}{He} + \mbox{D} &\rightarrow& \iso{5}{Li} + \gamma
\end{eqnarray}
but this is far slower than $p + {\rm D}$ reaction as well as
$\iso{5}{Li}$ being unstable to decay into $\iso{4}{He} + p$.  
This leaves only 
\begin{eqnarray}
\iso{3}{He} + \iso{4}{He} &\rightarrow& \iso{7}{Be} + \gamma
\end{eqnarray}
as the starting point into the proton-rich side of the isotopes since
$\iso{7}{Be}$ is stable in the Weakless Universe.
Proton-clumping interactions such as
\begin{eqnarray}
\iso{7}{Be} + p &\rightarrow& \iso{8}{B} + \gamma \\
\iso{8}{B} + p &\rightarrow& \iso{9}{C} + \gamma
\end{eqnarray}
can push up to $\iso{9}{C}$, but no further, since $\iso{10}{N}$ 
is not stable.  Helium-clumping interactions such as
\begin{eqnarray}
\iso{7}{Be} + \iso{3}{He} &\rightarrow& \iso{10}{C} + \gamma \\
\iso{7}{Be} + \iso{4}{He} &\rightarrow& \iso{11}{C} + \gamma \\
\iso{8}{B} + \iso{3}{He} &\rightarrow& \iso{11}{N} + \gamma \\
\iso{8}{B} + \iso{4}{He} &\rightarrow& \iso{12}{N} + \gamma 
\end{eqnarray}
also produce proton-rich isotopes of carbon and nitrogen.  
Further proton-clumping and helium-clumping can populate 
proton-rich isotopes of oxygen and higher.  The extent to which 
this series of reactions or the usual triple-$\alpha$ process 
dominates to generate carbon and other higher $Z$ elements 
depends on nuclear cross-sections, 
core temperatures, and detailed modeling of stellar cores.  
It would be extremely interesting to simulate stellar fusion given 
the initial abundances of protons, deuterium, and helium
in the absence of weak interactions.  This is, however, beyond 
the scope of this paper.

\section{Stellar Lifetimes}
\label{lifetime-sec}

A star is a self-controlled reactor in steady-state equilibrium. 
Its lifetime, once in steady-state, does not depend directly on the 
rates of micro-physical processes but rather on its gross 
thermodynamical properties.  This means that even though 
stars in the Weakless Universe begin fusion by the fast reaction
(\ref{D+p}), stellar lifetimes can nevertheless be billions of 
years long.

Steady-state equilibrium implies that the energy that is radiated away 
from the star's surface must equal the energy produced in its interior. 
This internal energy production comes from gravitational collapse 
and nuclear reactions.  Starting with a gas cloud of mass 
$M$ and a certain chemical composition, steady-state equilibrium may be 
stated as an equality of luminosities
\begin{eqnarray}  
L_\star(T,M,\ldots) &=& 
L_{\mathrm{nuc}}(T,M,\ldots) + L_{\mathrm{col}}(T,M,\ldots)
\label{theq}
\end{eqnarray}
where $L_\star$ is the stellar luminosity, $L_{\mathrm{nuc}}$ is the rate of
nuclear energy production and $L_{\mathrm{col}}$ is that from gravitational
collapse.  

The luminosity of a star, $L_\star$, is set by its large scale properties 
that are determined by hydrostatic and thermal equilibrium. 
Hydrostatic equilibrium requires that gravitational pressure must 
balance against a combination of thermal gas pressure, electron 
degeneracy pressure, and radiation pressure.  In the stars considered
here, as is the case in our Sun, the dominant pressure is the gas 
pressure which can be approximated by the ideal gas law, $P
\propto \rho T$. Hydrostatic equilibrium leads to a decreasing
temperature gradient from the stellar core to the surface. 

The details of the temperature gradient are also determined by 
the mechanism and rate of energy transport from the core.  
If the opacity is low enough, the dominant mechanism to 
transport energy from the core to the surface is by radiation,
as occurs for example for the Sun.  However, during 
the protostellar phase, when the Sun was much younger and colder, 
the opacity was much higher.  This inability to efficiently 
transport energy by radiation leads to an instability in the
pressure gradient of the star, causing convection to be the dominant 
energy transport mechanism.  Stars whose energy transport mechanism
is convection are said to be in the Hayashi phase.  
Convection is generally a more efficient transport mechanism, 
and so stars in their Hayashi phase are more luminous than they are 
in their later radiative phase\footnote{We thank Savas Dimopoulos and 
Jay Wacker for discussions about the Hayashi phase.}.

\begin{figure}
\begin{center}
\includegraphics[width=0.8\hsize]{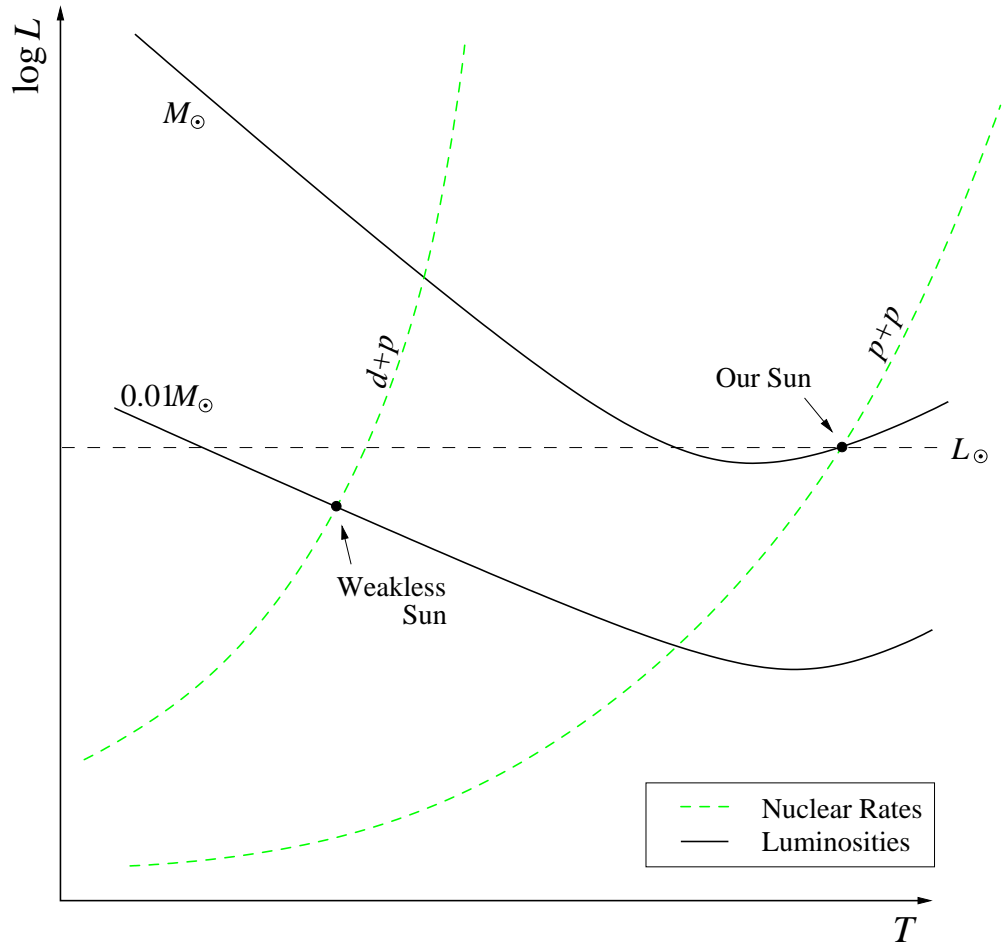}
\end{center}
\caption{A schematic plot of the stellar luminosity (black) and nuclear
burning rate (red) on a log scale as a function to the stellar core
temperature. The phase of decreasing luminosity is the Hayashi phase, when
convection is dominant. Stars will remain in a steady state when the
luminosity and nuclear rate are equal. The lifetime of the star in steady
state is the ratio of burnable energy to the steady state luminosity.
In the Weakless Universe one can get
long lives stars if the primordial deuterium abundance is higher than in our
universe, increasing the amount of burnable fuel. The longest living stars,
i.e.\ with low luminosity, are of order 0.01-0.02 $M_\odot$.}
\label{stars-fig}
\end{figure}

Given a detailed model for the equation of state and energy transport in
the star one can calculate the stellar radius and temperature profile
(generally using a numerical simulation).
The stellar luminosity $L_\star$ is given by black-body radiation
\begin{eqnarray}
L_\star &=& (4\pi R^2) \sigma T_{\rm eff}^4
\end{eqnarray}
where $\sigma$ is the Stefan-Boltzmann constant and $T_{\rm eff}$ is roughly
given by the surface temperature of the star.  To gain some appreciation
for the behavior of $L_\star$ with time, we can follow the evolutionary 
track of a solar mass star~\cite{Hayashi:1966tn} as is described 
schematically in Fig.~\ref{stars-fig}.  
A young star in the convective Hayashi phase is
initially much more luminous than our Sun, however its luminosity 
decreases with time.  Then the central temperature increases but the 
surface temperature remains roughly constant.  The decreasing radius 
leads to a drop in luminosity.  At some point radiation becomes the 
dominant energy transport mechanism and the surface temperature begins 
to rise as the star continues to collapse and the luminosity begins 
to climb with increasing core temperature.  Stars with smaller masses 
follow a roughly parallel track, but with a smaller luminosity
compared to a solar mass star.  However, the onset of the radiative
phase is delayed for stars with lower masses. 

Knowing at what rate a star radiates energy we now analyze the 
internal energy sources, i.e., the right hand side of Eq.~(\ref{theq}). 
The nuclear reaction rate is set by the details of the nuclear process 
in question.  Basically, it is set by the tunneling probability through 
a Coulomb barrier at a given temperature.  Any such rate will be a 
steeply climbing function of temperature.  At any given temperature, 
reactions involving the strong force will be much more rapid than those
involving the weak force, however this does not mean the star will be 
short-lived.  Instead, this allows the star to burn at a lower temperature.

When the star is young and cold, the nuclear reaction rates are 
exceedingly small.  The only source of energy satisfying Eq.~(\ref{theq}) 
is from gravitational collapse.  As the star contracts, half of the
gravitational potential energy is converted into heat, the other
half radiated away, satisfying the virial theorem. 
As the star continues to collapse, the central temperature increases,
the nuclear reaction rates rapidly increase, supplying a larger 
fraction of energy to the star.  This slows the gravitational collapse, 
maintaining thermal equilibrium.  The gravitational collapse 
halts once nuclear reactions saturate the stellar luminosity. 
When this occurs, a star is said to be in a steady-state that
defines the main sequence of stellar populations.  A star on the 
main sequence continues to shine without changing its gross features 
until it consumes its burnable nuclear material.  The lifetime of a star 
in steady state is thus
\begin{eqnarray}
t_\star &=& 
\frac{\mbox{Burnable Energy}}{\mbox{Luminosity at Steady State}} \; .
\end{eqnarray}
In the Weakless Universe there are two classes of long-lived stars:
deuterium burning and $\iso{3}{He}$ burning.

Consider first stars that burn deuterium.  In our Universe,
it is believed that brown dwarfs, which are stars in the range 
$\sim 0.01 - 0.2\ M_\odot$, burn deuterium for millions of years.
Due to the low primordial deuterium abundance of our Universe 
these stars never reach a deuterium burning steady-state 
\cite{burrows,Chabrier:1997vx}. In the Weakless Universe, 
the deuterium abundance is much higher and thus a deuterium burning 
star can easily achieve steady-state equilibrium.  
Detailed simulations of deuterium burning in brown dwarfs has been 
done for our Universe \cite{burrows}.
This simulation was repeated by Burrows \cite{burrows-private}
for the Weakless Universe with a primordial deuterium abundance 
of $y_d\sim 0.1$.\footnote{We are grateful to Adam Burrows for 
performing this simulation.}  
For example, a star with a mass of $0.02 M_\odot$ burned deuterium 
in steady-state for roughly 7 billion years with a luminosity of 
$0.025 L_\odot$.  Less massive stars burned for a longer 
time.\footnote{For simplicity, this simulation was performed 
at zero metallicity, but increasing the metallicity is only expected 
to modestly lower the mass of stars that burn deuterium 
for a long time.}

The second class of stars, which are of order solar mass, 
burn $\iso{3}{He}$ in steady-state.  These stars go through
the deuterium burning phase rapidly, converting essentially
all deuterium into $\iso{3}{He}$.  These star will then contract 
and heat up, eventually igniting $\iso{3}{He}$ through the fusion 
reaction Eq.~(\ref{He3-eq}).  This reaction provides over half
of the energy release from the Sun, and yet it is far slower than 
deuterium fusion due to the larger Coulomb barrier.  We expect
that $\iso{3}{He}$ burning to not begin in earnest until the star 
has collapsed nearly into its steady-state size and temperature,
leading to a main sequence star that will burn for billions 
of years.

\section{Supernovae and Populating the ISM}
\label{SN-sec}

In our Universe, supernovae are vital to provide
a mechanism to disperse heavy elements created inside a 
star out into the interstellar medium (ISM) that can later
gravitationally contract to form second-generation stars and planets.  
The final stages
of stellar evolution are therefore important to understand
in the Weakless Universe in order to determine exactly what happens 
to dying stars and consequently to what extent the interstellar medium 
can be populated with elements heavier than those created
during BBN\@.

In our Universe, stars slightly heavier than our Sun end in
supernovae whose explosions release heavy elements into 
the interstellar medium.  There are two main classes of supernovae:  
core-collapse
(type Ib, Ic, II) and accretion (type Ia).  In very massive
stars, the core increasingly becomes unfusable iron
supported by electron degeneracy pressure.  Once the 
core temperature of a very massive star gets high enough, 
electron degeneracy pressure fails due to the electrons
becoming relativistic.  This is the Chandrasekhar bound
for the core of the star, 1.4 $M_\odot$.  Electron degeneracy
pressure fails at 
$T \simeq m_e \sim 0.5$ MeV, which is coincidentally 
nearly the same temperature needed to have electrons 
recombine with protons to form neutrons, i.e., inverse $\beta$-decay
(also known as the Urca process in astrophysics).  

In the Weakless Universe, the electron degeneracy pressure of the core
does indeed fail for temperatures beyond $T \sim m_e$, but obviously 
inverse $\beta$-decay is absent.  This means that stellar cores of 
very massive stars will collapse and, for a mass range similar to 
our Universe, the collapse will halt when the density reaches 
nuclear density.  The collapse releases a huge amount of energy, 
and the abrupt halt at nuclear density leads to a shock wave, 
just as in our Universe.  Unlike our Universe, however,
the core has no mechanism to cool since there is no neutrino
emission.  This leaves the core at high temperature and pressure
that can support the outer layers of the star, leading to 
core-collapse supernova that ``fizzle''.\footnote{If axions 
were present in the 
Weakless Universe with a sufficiently large coupling, they could
lead to cooling of the supernova core that could be tuned to be 
equivalent to the cooling resulting from neutrinos.}
Notice also that final stage of these stars leads either to a
black hole or a stable nuclear-density object that contains
roughly equal numbers of protons and neutrons at nuclear density.
This is a rather unusual astrophysical object!\footnote{We thank
D.~Anderson and M.~Sher for discussions on this point.}

Accretion supernovae, however, lead to explosions in the 
Weakless Universe.  Here one imagines a star with a mass
that is slightly below the Chandrasekhar limit, but acquires
mass through accretion with, say, a binary companion. 
These white-dwarf stars have burned up to about carbon
and oxygen, and are marginally stable due to electron
degeneracy pressure.  Once accretion pushes the mass of these
stars beyond the Chandrasekhar limit, gravity wins over
electron degeneracy, and these stars collapse in a huge fusion
explosion with no significant fraction of energy released into neutrinos.
Again, the details of the explosion are quite different
in the Weakless Universe versus our Universe
because of the absence of the Urca process and its
effects on rates of nuclear chains that occur during the 
explosion.

The Weakless Universe therefore appears to have one explosive 
supernova mechanism for populating the ISM with elements well 
above helium.
There are also less violent processes, such as novae, 
which lead to large ejections of the outer layers of stars
thought to arise from binary systems in which the two
stars orbit each other very closely.  Each nova can release
as much as 0.1\% of the mass of a star, and so while
the mass release pales in comparison to a supernova, the 
far greater number of novae allow this process to potentially
populate the ISM with at least some of the products of 
stellar nucleosynthesis.

\section{A Natural Value of the Cosmological Constant?}

We have shown that even with electroweak breaking at the Planck scale, a
habitable universe can result so long as we are able to adjust technically
natural parameters.  It would be interesting to perform the same procedure 
for the cosmological constant (CC).  Here our goal is far more 
modest than in our previous discussion:  
we simply wish to examine whether large scale structure
and complex macroscopic systems can result if the cosmological constant is
pushed to the Planck scale while we freely adjust other parameters.
Performing a thorough analysis of this question is beyond the 
scope of this work.  We will instead simply sketch some of the 
issues by examining simplified toy models.  We find an upper bound 
on the CC from two qualitative requirements:  that density perturbations 
grow, and that complex macroscopic systems consist of a large number 
of particles.

In order for structure to be formed in our Universe, a period of 
matter domination is vital to allow for linear growth of 
density perturbations.  Matter domination is cutoff by 
CC domination, which is just the Weinberg bound on the 
anthropic size of the cosmological constant.  Naively we could
raise $\delta \rho/\rho$ up to order one, so that the bound on 
the cosmological constant relaxes to
\begin{eqnarray}
\rho_\Lambda &\lsim& T_{\rm eq}^4 \; . \label{matt-rad-bound}
\end{eqnarray}
But even this modest gain (about 10 orders of magnitude of 120)
is much too optimistic.  For Universes qualitatively similar
to ours, Refs.~\cite{Tegmark:1997in,Tegmark:2005dy} found 
other astrophysical constraints limit the size of density 
perturbations (and place constraints on other parameters,
such as the baryon density) such that the largest relaxation of 
the CC is closer to about 3 orders of magnitude of 120.  
Hence, even varying multiple cosmological parameters 
simultaneously, this appears to be as far as one can go without 
radically changing the Standard Model itself.

Suppose we generalize the Standard Model to an arbitrary
effective theory with matter and radiation.
Could the bound on the cosmological constant be further relaxed 
by adjusting the temperature of matter-radiation equality
in this generalized theory?  Consider a toy model in which
the matter is characterized by a single scale $\mu$ that sets 
the mass of the particles, given by
\begin{eqnarray}
\mu &\equiv& \epsilon \Mpl \; ,
\end{eqnarray}
where $\epsilon$ is sufficiently small so that we can analyze
this theory perturbatively.
The weakest bound on the cosmological constant arises when
the matter asymmetry (analogous to $\eta_b$ in our Universe) 
is order one, so that $T_{\rm eq} \sim \mu$.
Hence, the bound given by Eq.~(\ref{matt-rad-bound}) can be easily
avoided by increasing $\mu$, causing matter domination to occur 
at an earlier epoch in this universe.

However, this is not the only bound on the vacuum energy. 
In particular, density perturbations on scales larger than the 
horizon do not grow.  The largest scale that will collapse in 
any universe is the scale that enters the horizon as the universe 
is becoming CC dominated.  Therefore, there is a maximal mass for 
any structure \cite{Hellerman:2005yi} that is given by 
Eq.~(\ref{hell-eq}), which we rewrite here as
\begin{eqnarray}
M^{\rm max} &\sim& \Mpl^3 \rho_\Lambda^{-1/2}
\end{eqnarray}
for $\delta_0 = 1$.  Given any field theory, there will be some 
minimal mass, $M_\star$, for interesting structure to form. 
For example, in our Universe, a star which is below a few percent 
of a solar mass does not ignite.\footnote{This is also true in the 
Weakless Universe.}  A universe in
which all of the structure is less than this critical mass 
is probably anthropically uninteresting.  We can express the
the critical mass $M_\star$ as
\begin{eqnarray}
M_\star &\equiv& N\mu \; ,
\end{eqnarray}
where $N$ itself is a complicated function of $\mu$.  
Requiring that $M_\star\lesssim M^{max}$ we obtain the following 
bound on the CC,
\begin{eqnarray}
\rho_\Lambda &\lesssim& \frac{\Mpl^6}{M_\star^2} 
                \; = \; \frac{\Mpl^6}{N^2 \mu^2} \; .
\label{struct-bound}
\end{eqnarray}
Regardless of its detailed dependence on the model, $N$ is the 
number of particles that make up a complex, macroscopically 
interesting dynamical system and is thus expected to be huge.  

In our Universe with $\delta \rho/\rho \sim 1$, the bound 
Eq.~(\ref{matt-rad-bound}) is still more restrictive than
Eq.~(\ref{struct-bound}) when applied to structures such as 
galaxies and stars.
For example, for $M_\star$ equal to the mass of a typical galaxy, 
the bound becomes roughly $10^{-100} \Mpl^4 = (1 \; {\rm keV})^4$.  
For $M_\star = M_\odot$, the bound on 
$\rho_\Lambda$ is $10^{-80} \Mpl^4 = (10 \; {\rm MeV})^4$.
Nevertheless, this is obviously far smaller than the 
natural value, $\Mpl^4$.

For the simple example under consideration we can combine
Eqs.~(\ref{matt-rad-bound}) and (\ref{struct-bound}) to obtain
\begin{eqnarray}
\rho_\Lambda &\lesssim& \mbox{min} 
\left\{ \epsilon^4, \frac{1}{\epsilon^2 N^2} \right\} \times \Mpl^4 \; .
\end{eqnarray}
We find that by assuming $N$ is very large, as expected for a macroscopic
system, and $\epsilon$ is small, the upper bound for the vacuum energy is
always parametrically smaller than $\Mpl^4$.  

Here we have assumed an extremely simple model that has just one scale,
$\mu$.  However, an essentially similar set of bounds was derived
in~\cite{Hellerman:2005yi} for our Universe that has several important scales
(i.e., proton mass, dark matter, etc.). Tough we cannot exclude the
possibility the models that violate this bound exist, we find it telling that
the Planck scale seems out of reach. Taking the limit $\Lambda \ra \Mpl$
forces us into $\epsilon \ra 1$ and thus $N \ra 1$.  It seems hard to imagine
that any habitable universe results from structures that consist of single
particles.

\section{Discussion}

In this paper we have constructed a Universe without weak interactions 
that undergoes BBN, matter domination, structure formation, 
star formation, long periods of stellar burning, stellar nucleosynthesis up
to iron, star destruction by supernovae, and and dispersal of heavy elements
into the interstellar medium.  These properties of the Weakless Universe were
shown by a detailed analysis that matched to our Universe as closely as
possible by arbitrarily adjusting Standard Model and cosmological parameters.
The Weakless Universe therefore provides a simple explicit counter-example to
anthropic selection of a small electroweak breaking scale, so long as
we are allowed to simultaneously adjust technically natural parameters
relative to our observed Universe.  As an aside, we are unaware of 
any obstruction to obtain a ``partial Weakless Universe''
in which $v < \tilde{v} < \Mpl$ while allowing analogous adjustments of 
technically natural parameters.

This hypothetical universe is a counter-example to anthropic selection 
of the electroweak scale in the context of an effective field theory, 
where we are free to imagine arbitrary adjustments in technically 
natural parameters.  An ultraviolet completion, however, may or may 
not permit these parameter adjustments, and as a result the 
Weakless Universe may or may not be ``accessible''.  
This requires detailed knowledge of the ensemble of universes 
that are predicted.  String theory indeed appears to contain 
a huge number of vacua, a ``landscape'' 
\cite{Bousso:2000xa,Kachru:2003aw,Susskind:2003kw,Douglas:2003um}, 
in which some parameters adjust from one vacuum to another.  
Furthermore, only a specific set of parameters vary in the
field theory landscapes considered in \cite{Arkani-Hamed:2005yv}.
In its most celebrated form, the string landscape provides a potential 
anthropic rationale for the size of the cosmological 
constant~\cite{Weinberg:1987dv}.  
However, reliable model-independent correlations between the size 
of the CC and other parameters is lacking, and so we have no way to know 
yet whether the variation of parameters discussed here is realized 
on the string landscape.

It is instructive to compare the qualitative result of 
simultaneously varying several parameters with respect to the CC 
and contrasting this with the electroweak scale.  
For a universe that is closely related to ours, 
Refs.~\cite{Tegmark:1997in,Tegmark:2005dy} showed that variations
of other cosmological parameters, including $\delta \rho/\rho$
and $\eta_b$, lead to roughly the same bound on the cosmological
constant as Weinberg's original argument.  In other words, 
Weinberg's prediction that the CC should be near to the
anthropic boundary is not significantly affected by variations 
of at least some other (cosmological) parameters.  
The electroweak scale, by contrast, can be ``relaxed'' 
all the way up to the Planck scale with the parameter 
adjustments that characterize the Weakless Universe.  
Even for a generalized universe characterized by matter with a single
mass scale, requiring large scale structure be made up of large 
numbers of matter particles is enough to show that the 
cosmological constant must be parametrically smaller than the 
Planck scale.  We conclude that the fine-tuning problems associated 
with the electroweak breaking scale and the cosmological constant 
appear to be qualitatively different from the perspective of obtaining
a habitable universe.

\section*{Acknowledgments}

We acknowledge numerous colleagues for valuable discussions, particularly 
D.~Anderson,
Z.~Chacko,
N.~Dalal,
S.~Dimopoulos,
S.~Hsu,
M.~Kamionkowski,
I.~Klebanov,
H.~Murayama,
M.~Papucci, 
M.~Peskin,
K.~Rajagopal,
E.~Ramirez-Ruiz,
M.~Schwartz, 
M.~Sher, 
B.~van Kolck, and
J.~Wacker.
We especially thank A.~Burrows for running his code with a high deuterium
abundance.  GDK and GP thank the Aspen Center for Physics, Stanford Linear
Accelerator Center, and the Institute of Theoretical Physics at Technion 
University for hospitality where part of this work was completed.  
GDK is grateful to Lawrence Berkeley Laboratory for hospitality where 
this work was initiated.  This work was supported in part
by DOE under contracts DE-AC02-76SF00515 (RH), DE-FG06-85ER40224 (GDK), and
DE-AC02-05CH11231 (GP).


\end{document}